# Simplified Approach to Implementing Controlled Unitary Operations in a Two-Qubit System

P. Kumar and S. R. Skinner

*Department of Electrical and Computer Engineering, Wichita State University, Wichita, Kansas 67260*

We introduce a scheme for realizing arbitrary controlled-unitary operations in a two qubit system. If the 2 × 2 unitary matrix is special unitary (has unit determinant), the controlled-unitary gate operation can be realized in a single pulse operation. The pulse, in our scheme, will constitute varying one of the parameters of the system between an arbitrarily maximum and a "calculated" minimum value. This parameter will constitute the variable parameter of the system while the other parameters, which include the coupling between the two qubits, will be treated as fixed parameters. The values of the parameters are what we solve for using our approach in order to realize an arbitrary controlled-unitary operation. We further show that the computational complexity of the operation is no greater than that required for a Controlled-NOT (CNOT) gate. Since conventional schemes realize a controlled-unitary operation by breaking it into a sequence of single-qubit and CNOT gate operations, our method is an improvement because we not only require lesser time duration, but also fewer control lines, to implement the same operation. To demonstrate improvement over other schemes, we show, as examples, how two controlled-unitary operations, one being the controlled-Hadamard gate, can be realized in a single pulse operation using our scheme. Furthermore, our method can be applied to a wide range of coupling schemes and can be used to realize gate operations between two qubits coupled via Ising, Heisenberg and anisotropic interactions.

## I. Introduction

In a traditional computer, information is encoded in bits and is manipulated using Boolean logic gates. Likewise, a quantum computer manipulates qubits by executing a series of quantum gates [1-4]. A qubit is a quantum two-level system where the two logic states 0 and 1 are represented by the basis states, represented as $|0\rangle$ and $|1\rangle$, of a two-dimensional Hilbert space and each single-qubit quantum gate is a unitary transformation acting on the qubit. Two qubit operations are controlled interactions between two qubits which produce coherent changes in the state of one qubit that are contingent upon the state of another. The Controlled-NOT (CNOT) gate is the prototypical two-qubit quantum logic gate. Under the operation of this gate, the target qubit flips its state if and only if the control qubit is in the $|1\rangle$ state.

In quantum computing, the CNOT gate and single-qubit gates are considered elementary gates and any n-qubit operation can be decomposed into a combination of these gates [1,5]. Recent work shows that the CNOT gate is also one of the most efficient quantum gates known, and that three CNOT gates are necessary and sufficient in order to implement an arbitrary unitary transformation of two qubits, supplemented with single qubit rotations [6-8]. In [9], S. Bullock showed how to decompose an arbitrary two-qubit computation into 23 elementary gates or less. To this end, he showed that a lower bound for worst-case optimal two-qubit circuits calls for at least 17 gates: 15 single-qubit rotations and 2 CNOTs. Since then much work has been done in devising methods for decomposing a desired quantum computation into the shortest possible sequence of one-qubit and two-qubit gates [7,8,10,11]. Several methods have been proposed for devising quantum circuits that are universal and can implement any arbitrary unitary operation [6,8,12-14]. Recently, it was shown that any interaction that can create entanglement between any pair of qubits is universal for quantum computation together with single-qubit gates [12, 13]. Following this, Zhang et al. developed an analytic method to construct exact quantum universal circuits given an arbitray entangling gate together with local gates[14].

Several important applications in quantum computing, including implementations of quantum algorithms like Shor's factorization [1,15], require controlled-unitary operations. Controlled-unitary operations are two-qubit operations where an operation is performed on the target qubit if and only if the control qubit is in the $|1\rangle$ state. When the control qubit is in the $|0\rangle$ state, no operation is performed on the target qubit. In [5], Barenco, *et al.*, showed that any two-qubit controlled-unitary operation can be achieved using six basic gates – four single-qubit and two CNOT gates. Using this result, Song and Klappenecker derived the minimal number of elementary gates necessary in any implementation of a controlled-unitary gate [16]. In this paper, we show a means of realizing a controlled-unitary operation, up to an overall global phase, in a *single pulse operation* instead of decomposing it into elementary gate operations. The pulse, in our scheme, will constitute varying one of the parameters of the system between an arbitrarily maximum and a "calculated" minimum value. Since the computational complexity of a task depends on how difficult it is to realize the unitary transformation, the question arises whether this single pulse operation as described by our scheme is easy to implement on a two-qubit system. The answer to this question depends upon the physical system under consideration. Considering superconducting quantum interference devices (SQUIDs) as an example [17-25], the pulse would constitute varying the bias acting on the target qubit, since this is relatively easy to control.

Previously we have showed how to implement a CNOT gate in a single pulse operation. We used a means of reducing the 4 × 4 Hamiltonian of the coupled system to a 2 × 2 Hamiltonian describing the evolution of the target qubit. Here, we extend the same approach towards realizing general controlled-unitary operations in **SU(2)**. [In Mathematics, the special unitary group of degree *n*, denoted by **SU(n)**, is the group of *n* × *n* matrices with unit determinant]. This allows the possibility of realizing controlled-unitary operations without having to break them down into CNOT and single qubit operations, which is desirable from a quantum computing architectural point of view, since each operation on a qubit opens up the system to a noisy environment and the number of gate operations must be kept minimal. We show that depending upon the physical system under consideration, the complexity of the task of realizing an arbitrary controlled-unitary operation in **SU(2)** is no more difficult than realizing a CNOT gate. In [26], we did not show the actual derivation of the 2 × 2 reduced Hamiltonian from the "unreduced" 4 × 4 coupled system Hamiltonian. We show this derivation here in Section III. Moreover, in [26], we did not compare the evolution of the system under the reduced Hamiltonian to that under the unreduced coupled system Hamiltonian. Here we make such a comparison and show that the evolution of the system under the reduced Hamiltonian is a close approximation to the evolution under the unreduced coupled system Hamiltonian. Furthermore, the reduced Hamiltonian technique used to realize a CNOT gate operation in [26] assumes an Ising type interaction. In this paper, we show that the scheme is not restricted to qubits coupled via the Ising interaction, but can be extended to systems where the two qubits interact through Heisenberg and anisotropic interactions, making our pulsed bias method more general. We also discuss and analyze some of the pitfalls of this technique in this paper.

Recently, Zhou *et al.* devised a scheme for universal and scalable quantum computation without the need to tune the couplings between qubits [27]. Their method relies on the idea of computing with logical qubits, which comprise several physical qubits [28]. The coupling between the encoded qubits are effectively turned on and off by computing in and out of carefully designed interaction free subspaces analogous to decoherence free subspaces [29, 30]. It is important to point out that while our scheme, like theirs, does not require the coupling to be treated as a variable parameter, the gate operations in our scheme are implemented on the states of physical qubits and not of encoded qubits.

## II. Single Qubit Evolution

The evolution of a single qubit system with basis states $|0\rangle$ and $|1\rangle$ is governed by a 2 × 2 Hamiltonian, $\mathbf{H_1}$, given as:

$$\mathbf{H_1} = \begin{pmatrix} \varepsilon & \Delta - ik \\ \Delta + ik & -\varepsilon \end{pmatrix} \tag{1}$$

Here, $\Delta$, $\varepsilon$, $k$ are parameters of the system under consideration. For instance, for superconducting qubits, $\Delta$ is the tunneling parameter and $\varepsilon$ is the bias acting on the qubit ($k=0$). Given an initial state for a single-qubit system, its state after some time duration "t" can be evaluated by integrating the Schrödinger wave equation. The state of a single qubit two-level system at any time "t" is thus related to its initial state by a unitary transformation, $\mathbf{W}$, which can be represented as a 2 × 2 matrix:

$$\mathbf{W} = \begin{pmatrix} \cos\left(2\pi\left(\sqrt{\Delta^2 + \varepsilon^2 + k^2}\right)t\right) - i\varepsilon \frac{\sin\left(2\pi\left(\sqrt{\Delta^2 + \varepsilon^2 + k^2}\right)t\right)}{\sqrt{\Delta^2 + \varepsilon^2 + k^2}} & (-i\Delta - k)\frac{\sin\left(2\pi\left(\sqrt{\Delta^2 + \varepsilon^2 + k^2}\right)t\right)}{\sqrt{\Delta^2 + \varepsilon^2 + k^2}} \\ (-i\Delta + k)\frac{\sin\left(2\pi\left(\sqrt{\Delta^2 + \varepsilon^2 + k^2}\right)t\right)}{\sqrt{\Delta^2 + \varepsilon^2 + k^2}} & \cos\left(2\pi\left(\sqrt{\Delta^2 + \varepsilon^2 + k^2}\right)t\right) + i\varepsilon \frac{\sin\left(2\pi\left(\sqrt{\Delta^2 + \varepsilon^2 + k^2}\right)t\right)}{\sqrt{\Delta^2 + \varepsilon^2 + k^2}} \end{pmatrix} \tag{2}$$

Note that the $\mathbf{W}$ matrix belongs to the Lie group **SU(2)**. Using Eq. (2), we can find the state of the qubit at any time "t" and the probability of the qubit in the $|1\rangle$ state can be written as an oscillatory function of time as:

$$P_{|1\rangle} = X \mp Y\cos(2\pi ft) \tag{3}$$

where the '-' and '+' are used when the qubit starts out in the $|0\rangle$ and $|1\rangle$ states, respectively, as its initial state. The offset $X$, the amplitude $Y$ and the frequency $f$ of probability oscillation are given as:

$$\text{Offset, } X = \frac{1}{2} \mp \frac{\varepsilon^2}{2(\Delta^2 + k^2 + \varepsilon^2)}, \tag{4}$$

$$\text{Amplitude, } Y = \frac{\Delta^2 + k^2}{2(\Delta^2 + k^2 + \varepsilon^2)}, \tag{5}$$

$$\text{Frequency, } f = 2\sqrt{\left(\Delta^2 + k^2 + \varepsilon^2\right)}, \tag{6}$$

Here, we have used units in which the Planck's constant is 1.

When $\varepsilon$ is much larger than $\Delta$ and $k$, the qubit can be maintained in the state it has been initialized to. Therefore, in the limit as $\varepsilon \to \infty$, the amplitude of oscillations of the probability function given by (3) reduce to zero whereby the qubit maintains its state. In this case, the probability of the qubit in the $|1\rangle$ state is given as:

$$P_{|1\rangle} = X = \frac{1}{2} \mp \frac{\varepsilon^2}{2\left(\Delta^2 + \varepsilon^2 + k^2\right)} \approx \frac{1}{2} \mp \frac{\varepsilon^2}{2\varepsilon^2} = \frac{1}{2} \mp \frac{1}{2} \tag{7}$$

which evaluates to 0 if the qubit starts out in the $|0\rangle$ state, and to 1 if it starts out in the $|1\rangle$ state. Likewise, if the qubit starts out in any superposition of the two basis states, it remains in the state. Therefore, when $\varepsilon \gg \Delta$, we can think of $\Delta$ as having no effect on the evolution of the qubit under the influence of $\varepsilon$. We can, thus, safely omit the contribution of $\Delta$ to the Hamiltonian described in Eq. (1) and instead write the Hamiltonian as:

$$\mathbf{H}_1 \approx \begin{pmatrix} \varepsilon & 0 \\ 0 & -\varepsilon \end{pmatrix} \tag{8}$$

On the contrary, when $\varepsilon$ is zero, the qubit undergoes pure oscillations. The probability of the qubit in the $|1\rangle$ state is given as:

$$P_{|1\rangle} = X = \frac{1}{2} \mp \frac{\Delta^2 + k^2}{2\left(\Delta^2 + k^2\right)} \cos\left(2\pi\sqrt{\left(\Delta^2 + k^2 + \varepsilon^2\right)}t\right) = \frac{1 \mp \cos\left(2\pi\sqrt{\left(\Delta^2 + k^2 + \varepsilon^2\right)}t\right)}{2} \tag{9}$$

which is a purely oscillatory function with frequency of oscillation given by Eq. (6). By controlling the time duration of the oscillations for a chosen value of $\Delta$, single qubit rotations are thus realized.

### III. Coupled System Evolution – Reduced Hamiltonian

Consider a system of two qubits, A and B, which interact with each other via an Ising interaction. The Ising type coupling between qubits is commonly seen in proposals for superconducting Josephson junction qubits [30, 31] and also arises as one limit of dipole-dipole or J-coupling systems [30, 32]. The Hamiltonian describing the evolution of this system is a 4 × 4 matrix:

$$\mathbf{H}_2 = \begin{pmatrix} \varepsilon_A + \varepsilon_B + \xi & \Delta_B - ik_B & \Delta_A - ik_A & 0 \\ \Delta_B + ik_B & \varepsilon_A - \varepsilon_B - \xi & 0 & \Delta_A - ik_A \\ \Delta_A + ik_A & 0 & -\varepsilon_A + \varepsilon_B - \xi & \Delta_B - ik_B \\ 0 & \Delta_A + ik_A & \Delta_B + ik_B & -\varepsilon_A - \varepsilon_B + \xi \end{pmatrix} \tag{10}$$

Here $\Delta_A$, $\Delta_B$, $\varepsilon_A$, $\varepsilon_B$, $k_A$ and $k_B$ are the parameters for qubits $A$ and $B$, and $\xi$ is the coupling constant between the qubits. Suppose we want to perform a unitary operation on qubit B only, without performing any operation on qubit A. Since qubit A is coupled to qubit B, the evolution of qubit B will depend on the state of qubit A. In order to isolate qubit B from qubit A, we need to switch off the coupling between the two qubits, whereby we can treat qubit B as a single qubit system governed by a Hamiltonian of the form of Eq. (1). However, the ability to tune the coupling requires an additional control line and it might be more desirable to perform single qubit operations on qubit B without having to switch off the coupling. In practical quantum computing applications, varying the coupling between qubits might not always be possible. For instance, in Josephson junction devices [17, 18, 30, 35-39], the coupling is usually realized using a hard-wired capacitor or inductor which is fixed during fabrication and therefore, cannot be tuned during computation. Even though a number of variable coupling schemes [17, 29, 37, 39] have been devised, they are not completely satisfactory. Most of them require external controls making them major decoherence sources [37, 39], while others avoiding the use of external controls in their design are limited in the number of qubits that can be incorporated [17, 30]. Therefore, a scheme which allows implementation of controlled-unitary operations without switching the couplings is very useful because, besides reducing decoherence, it simplifies the operation drastically. Here we describe how the 4 × 4 Hamiltonian describing the two-qubit system can be reduced to a single qubit system of the form of Eq. (1) describing the evolution of qubit B only, in the presence of the coupling term.

To this, we take a closer look at the Hamiltonian described by Eq. (10). Suppose the parameter $\varepsilon_A$ is made much higher than the parameter $\Delta_A$ wherein the effect of the latter as compared to the former on the evolution of qubit A will be negligible and thus, the probability of qubit A remaining in a state it has been initialized to will be close to 1. This is similar to what happens in a single qubit system as described by Eq. (8). Therefore, in the same way as we derived Eq. (8) from Eq. (1) by assuming $\Delta$ to be zero, we can safely assume $\Delta_A$ to be zero in Eq. (10) whereby the matrix given by (10) now has the following form:

$$\mathbf{H}_2 \approx \begin{pmatrix} \varepsilon_A + \varepsilon_B + \xi & \Delta_B - ik_B & 0 & 0 \\ \Delta_B + ik_B & \varepsilon_A - \varepsilon_B - \xi & 0 & 0 \\ 0 & 0 & -\varepsilon_A + \varepsilon_B - \xi & \Delta_B - ik_B \\ 0 & 0 & \Delta_B + ik_B & -\varepsilon_A - \varepsilon_B + \xi \end{pmatrix} \quad (11)$$

As can be seen from Eq. (11), the matrix is comprised of two blocks of 2 × 2 matrices which represent the two subspaces of qubit B in the four-dimensional Hilbert space depending upon the state of qubit A. Notice that the bias term, $\varepsilon_A$, only corresponds to a shift in energy in each subspace. In other words, the two energy levels |00⟩ and |01⟩ are centered on "$\varepsilon_A$" and the two energy levels |10⟩ and |11⟩ are centered on "-$\varepsilon_A$". Since this only contributes to an overall global phase factor in the evolution, it can be ignored. Therefore, the evolution of qubit B can be described by two single-qubit Hamiltonians of the form of Eq. (1) as follows:

$$\mathbf{H}_{B1} = \begin{pmatrix} \varepsilon_B + \xi & \Delta_B - ik_B \\ \Delta_B + ik_B & -(\varepsilon_B + \xi) \end{pmatrix} \quad (12)$$

$$\mathbf{H}_{B2} = \begin{pmatrix} \varepsilon_B - \xi & \Delta_B - ik_B \\ \Delta_B + ik_B & -(\varepsilon_B - \xi) \end{pmatrix} \quad (13)$$

where $\mathbf{H}_{B1}$ and $\mathbf{H}_{B2}$ are the corresponding matrices in the subspaces where qubit A is in the |0⟩ and |1⟩ states, respectively. The coupling term, $\xi$, adds to the parameter $\varepsilon_B$ in the subspace where qubit A is in the |0⟩ state and it subtracts from $\varepsilon_B$ in the subspace where qubit A is in the |1⟩ state. This is because the expectation values of the $\sigma_{ZA}$ operator are +1 and -1 in the subspaces where the states of qubit A are |0⟩ and |1⟩, respectively.

Given that qubit B starts off in some initial state |ψ(0)⟩, its state at some time "t" is given by integrating the Schrödinger equation as follows:

$$|\psi(t)\rangle = \exp(-i 2\pi t H_B) |\psi(0)\rangle \quad (14)$$

where $\mathbf{H}_B$ is one of the two matrices given by Eqs. (12) and (13) depending on whether qubit A is in the |0⟩ or |1⟩ state. Note that since we are dealing with subspaces of qubit B, |ψ(0)⟩ represents a two-dimensional vector.

Since the reduced Hamiltonian approach described by us is based on an approximation, we next show the effect of this approximation on the solution by comparing the evolution of a two-qubit system using the unreduced (exact) and reduced Hamiltonians. As an example, suppose the system initially starts out in the partially entangled state:

$$\frac{\sqrt{3}}{2}|00\rangle + \frac{\sqrt{3}}{4}|10\rangle + \frac{1}{4}|11\rangle \quad (15)$$

We choose the values of the parameters as follows: $\Delta_A = \Delta_B = 50$ MHz, $\varepsilon_A = 10$ GHz, $\varepsilon_B = 30$ MHz, $\xi = 12.5$ MHz, and t = 10 ns These values are experimentally realizable for SQUID experiments [26]. Notice that $\varepsilon_A$ is much higher than $\Delta_A$ since we desire to freeze the dynamics of qubit A. We next simulate the evolution of the two-qubit coupled system under the 4 × 4 Hamiltonian as given by Eq. (10) using the chosen parameters. Figure 1 shows the evolution of the probabilities of the four basis states. The final state of the system is:

$$0.45e^{i135°}|00\rangle + 0.30e^{i90°}|01\rangle + 0.19e^{i171°}|10\rangle + 0.06e^{i166°}|11\rangle \quad (16)$$

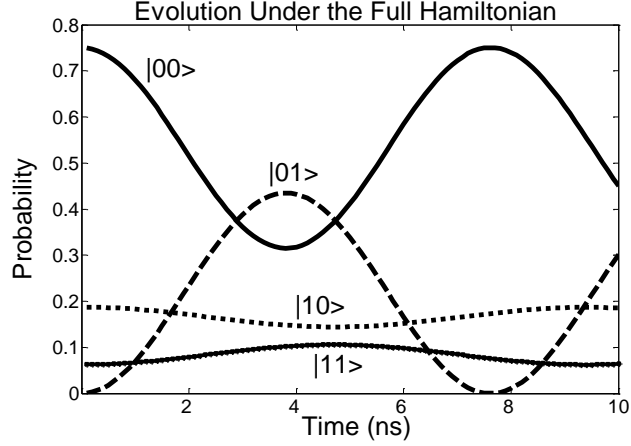

Fig. 1. Evolution of the probabilities of the 4 basis states of the 2-qubit system starting in the initial state given by Eq. (15).

We next study the evolution of the system under the reduced Hamiltonian approach. From Eq. (15) we can see that the system of two qubits is in a partially entangled state. The question arises as to what the initial state of qubit B will be in each of the two subspaces given this initial state. From Eq. (15), we can see that the probability amplitudes in the states $|00\rangle$, $|10\rangle$ and $|11\rangle$ are $\sqrt{3}/2$, $\sqrt{3}/4$, and $1/4$, respectively. Therefore, we can conclude that the initial state of B is $\sqrt{3}/2\,|0\rangle$ and $\sqrt{3}/4\,|0\rangle + 1/4\,|1\rangle$ in the subspaces where the state of qubit A is $|0\rangle$ and $|1\rangle$, respectively. Notice that we cannot use the term "qubit" to address B in each subspace since the state of B is not normalized.

We now simulate the evolution of B under the Hamiltonian $\mathbf{H_{B1}}$ with B starting out in the initial state $\sqrt{3}/2\,|0\rangle$. Figure 2 shows the probabilities of the $|0\rangle$ and $|1\rangle$ states which are identical to the plots for the $|00\rangle$ and $|01\rangle$ states in Fig. 1. The final state of B is:

$$0.45e^{i137°}|0\rangle + 0.30e^{i90°}|1\rangle. \quad (17)$$

Next, we simulate the evolution of B under the Hamiltonian $\mathbf{H_{B2}}$ with B starting out in the initial state $\sqrt{3}/4\,|0\rangle + 1/4\,|1\rangle\rangle$. Figure 3 shows the probabilities of the $|0\rangle$ and $|1\rangle$ states which are identical to the plots for the $|10\rangle$ and $|11\rangle$ states in Fig. 1. The final state of B under this evolution is:

$$0.19e^{i171°}|0\rangle + 0.06e^{i166°}|1\rangle. \quad (18)$$

Equations (17) and (18) give the final state of B in each subspace. While we might be tempted to think that the overall final state of B is a superposition of these two states (notice that Eqs. (17) and (18) together satisfy the normalization condition for a qubit state), we have to keep in mind that the evolution of qubit B is taking place in two different subspaces, and therefore, we cannot combine the two states directly to give a single superposition state for qubit B. However, we can now write the overall state of the two-qubit system by including the corresponding states of qubit A in each subspace. That is, using Eqs. (17) and (18), the final state of the two-qubit system can be written as:

$$|0\rangle(0.45e^{i137°}|0\rangle + 0.30e^{i90°}|1\rangle) + |1\rangle(0.19e^{i171°}|0\rangle + 0.06e^{i166°}|1\rangle) \quad (19)$$

which is of the same form as (16).

We can, therefore, see that the reduced Hamiltonian approach can be used to approximate the evolution of a two-qubit system with high accuracy when the state of one of the qubits remains frozen during the evolution. Since in controlled-unitary operations, the control qubit never changes its state, this approach can therefore be extended towards realizing arbitrary controlled-unitary operations when the 2 × 2 unitary matrix belongs to **SU(2)**. This is because the unitary matrices generated by Hamiltonians $H_{B1}$ and $H_{B2}$ are of the form of Eq. (2) and are therefore, special unitary. In the following section, we will show how such two-qubit gate operations can be achieved.

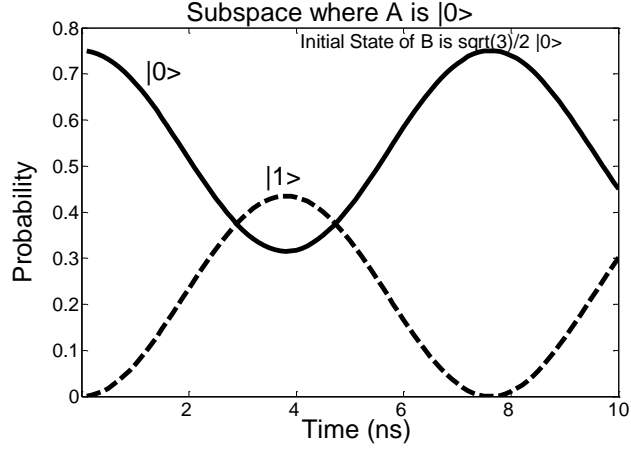

Fig. 2. Evolution of the probabilities of the states |0⟩ and |1⟩ of B in the subspace governed by Hamiltonian $H_{B1}$ where qubit A is in the |0⟩ state. B starts out in the initial state |0⟩ with a probability amplitude √(3)/2. The plots for the evolution of the |0⟩ and |1⟩ states are identical to the plots for the |00⟩ and |01⟩ states shown in Fig. 1.

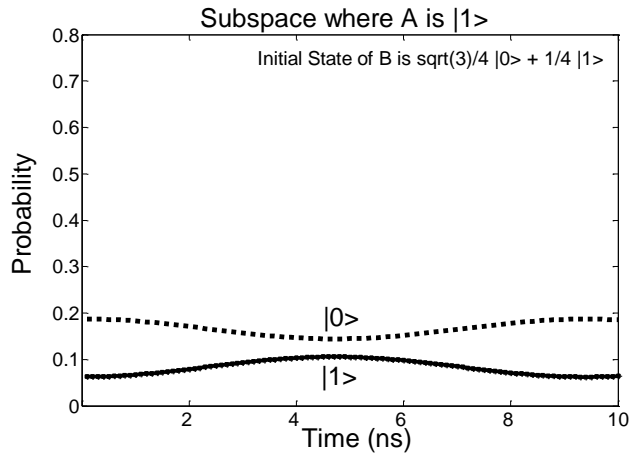

Fig. 3. Evolution of the probabilities of the states |0⟩ and |1⟩ of B in the subspace governed by Hamiltonian $H_{B2}$ where qubit A is in the |1⟩ state. B starts out in the initial state √(3)/4 |0⟩ + ¼ |1⟩. The plots for the evolution of the |0⟩ and |1⟩ states are identical to the plots for the |10⟩ and |11⟩ states in Fig. 1.

## IV. Realizing Controlled-Unitary Operations

It is a well known result that every 2 × 2 unitary matrix, **U**, belonging to **SU(2)** can be expressed as a product of three rotation matrices, two about the z axis (by angles β and δ) and one about the y axis (by an angle γ), as follows [1,5]:

$$\mathbf{U} = \begin{pmatrix} \exp\left(-i\left(\frac{\beta+\delta}{2}\right)\right)\cos\left(\frac{\gamma}{2}\right) & -\exp\left(-i\left(\frac{\beta-\delta}{2}\right)\right)\sin\left(\frac{\gamma}{2}\right) \\ \exp\left(+i\left(\frac{\beta-\delta}{2}\right)\right)\sin\left(\frac{\gamma}{2}\right) & \exp\left(+i\left(\frac{\beta+\delta}{2}\right)\right)\cos\left(\frac{\gamma}{2}\right) \end{pmatrix} \qquad (20)$$

where the values of β, δ and γ can be chosen to realize the desired unitary transformation. Therefore, for a chosen time step $T$, the parameters ($\Delta$ and $\varepsilon$) of a single-qubit system can be solved for by equating matrices (2) and (20) such that the **W** matrix realizes the unitary transformation represented by the **U** matrix.

Since under a controlled-unitary operation in a two-qubit system, we do not require the control qubit, A, to change its state, by making $\varepsilon_A$ large, we can force it to remain in its initialized state during the entire gate operation. This allows us to reduce the 4 × 4 Hamiltonian of the two-qubit system to a 2 × 2 Hamiltonian describing the evolution of the target qubit B as given by Eqs. (12) and (13). We can, therefore, write unitary matrices describing the evolution of the target qubit in the form of Eq. (2) under each of the two reduced Hamiltonians:

$$\mathbf{W}_{B1} = \begin{pmatrix} \cos\left(2\pi\sqrt{\Delta^2+(\varepsilon+\xi)^2+k^2}\,t\right) - i(\varepsilon+\xi)\dfrac{\sin\left(2\pi\sqrt{\Delta^2+(\varepsilon+\xi)^2+k^2}\,t\right)}{\sqrt{\Delta^2+(\varepsilon+\xi)^2+k^2}} & (-k-i\Delta)\dfrac{\sin\left(2\pi\sqrt{\Delta^2+(\varepsilon+\xi)^2+k^2}\,t\right)}{\sqrt{\Delta^2+(\varepsilon+\xi)^2+k^2}} \\ (k-i\Delta)\dfrac{\sin\left(2\pi\sqrt{\Delta^2+(\varepsilon+\xi)^2+k^2}\,t\right)}{\sqrt{\Delta^2+(\varepsilon+\xi)^2+k^2}} & \cos\left(2\pi\sqrt{\Delta^2+(\varepsilon+\xi)^2+k^2}\,t\right) + i(\varepsilon+\xi)\dfrac{\sin\left(2\pi\sqrt{\Delta^2+(\varepsilon+\xi)^2+k^2}\,t\right)}{\sqrt{\Delta^2+(\varepsilon+\xi)^2+k^2}} \end{pmatrix}$$

(21)

$$\mathbf{W}_{B2} = \begin{pmatrix} \cos\left(2\pi\sqrt{\Delta^2+(\varepsilon-\xi)^2+k^2}\,t\right) - i(\varepsilon-\xi)\dfrac{\sin\left(2\pi\sqrt{\Delta^2+(\varepsilon-\xi)^2+k^2}\,t\right)}{\sqrt{\Delta^2+(\varepsilon-\xi)^2+k^2}} & (-k-i\Delta)\dfrac{\sin\left(2\pi\sqrt{\Delta^2+(\varepsilon-\xi)^2+k^2}\,t\right)}{\sqrt{\Delta^2+(\varepsilon-\xi)^2+k^2}} \\ (k-i\Delta)\dfrac{\sin\left(2\pi\sqrt{\Delta^2+(\varepsilon-\xi)^2+k^2}\,t\right)}{\sqrt{\Delta^2+(\varepsilon-\xi)^2+k^2}} & \cos\left(2\pi\sqrt{\Delta^2+(\varepsilon-\xi)^2+k^2}\,t\right) + i(\varepsilon-\xi)\dfrac{\sin\left(2\pi\sqrt{\Delta^2+(\varepsilon-\xi)^2+k^2}\,t\right)}{\sqrt{\Delta^2+(\varepsilon-\xi)^2+k^2}} \end{pmatrix}$$

(22)

Here, $\mathbf{W}_{B1}$ and $\mathbf{W}_{B2}$ are the corresponding unitary matrices in the subspaces of qubit B governed by Hamiltonians $\mathbf{H}_{B1}$ and $\mathbf{H}_{B2}$, respectively. Notice that we have dropped the subscript B from the parameters in these equations.

Under a controlled-unitary operation, when the control qubit is in the $|0\rangle$ state, we do not want to perform any operation on the target qubit. Since the reduced Hamiltonian, $\mathbf{H}_{B1}$, governs the evolution of the target qubit in this case, *we want the matrix $\mathbf{W}_{B1}$ to realize a 2 × 2 identity matrix*. This can be achieved by choosing the argument of the sine term to be an even integer multiple of π. (We choose an "even" integer multiple of π because this condition causes the diagonal terms of Eq. (24) to evaluate to 1. For an "odd" integer multiple of π, it would evaluate to -1). Therefore, the following condition needs to be satisfied:

$$\left(\sqrt{\Delta^2+k^2+(\varepsilon+\xi)^2}\right)T = P \tag{23}$$

where $P$ is an integer and $T$ is a chosen time step within which we want to achieve the gate operation. Equation (23) will always have to be satisfied for controlled-unitary operations because we always require that when the control qubit is in the $|0\rangle$ state, an identity transformation be performed on the target qubit.

Next, when the control qubit A is in the $|1\rangle$ state, we want to realize a unitary transformation **U** on the target qubit as given by Eq. (20) (Recall that for a given unitary transformation, the values of the angles, β, δ and γ are known). Since the reduced Hamiltonian $\mathbf{H}_{B2}$ governs the evolution of the target qubit in this case, *we want the matrix, $\mathbf{W}_{B2}$, to realize the **U** matrix*. This can be achieved by equating the entries of the matrices given by Eqs. (20) and (22) whereby we arrive at the following set of equations:

$$\cos\left(\frac{\beta+\delta}{2}\right)\cos\left(\frac{\gamma}{2}\right) = \cos\left(2\pi\sqrt{(\Delta^2+k^2+(\varepsilon-\xi)^2)}\,T\right) \tag{24}$$

$$\sin\left(\frac{\beta+\delta}{2}\right)\cos\left(\frac{\gamma}{2}\right) = \frac{\varepsilon-\xi}{\sqrt{\left(\Delta^2+k^2+(\varepsilon-\xi)^2\right)}}\sin\left(2\pi\sqrt{\left(\Delta^2+k^2+(\varepsilon-\xi)^2\right)}\,T\right) \quad (25)$$

$$\cos\left(\frac{\beta-\delta}{2}\right)\sin\left(\frac{\gamma}{2}\right) = \frac{k}{\sqrt{\left(\Delta^2+k^2+(\varepsilon-\xi)^2\right)}}\sin\left(2\pi\sqrt{\left(\Delta^2+k^2+(\varepsilon-\xi)^2\right)}\,T\right) \quad (26)$$

$$\sin\left(\frac{\beta-\delta}{2}\right)\sin\left(\frac{\gamma}{2}\right) = \frac{-\Delta}{\sqrt{\left(\Delta^2+k^2+(\varepsilon-\xi)^2\right)}}\sin\left(2\pi\sqrt{\left(\Delta^2+k^2+(\varepsilon-\xi)^2\right)}\,T\right) \quad (27)$$

Note that the time step $T$ used in Eqs. (24), (25), (26) and (27) is the same time step used in Eq. (23).

Next, by dividing Eq. (27) by Eq. (26), we get the following relation:

$$-\frac{\Delta}{k} = \tan\left(\frac{\beta-\delta}{2}\right) \quad (28)$$

Dividing Eq. (27) by Eq. (25), we get the following relation:

$$-\frac{\Delta}{\varepsilon-\xi} = \frac{\tan\left(\frac{\gamma}{2}\right)\sin\left(\frac{\beta-\delta}{2}\right)}{\sin\left(\frac{\beta+\delta}{2}\right)} \quad (29)$$

From Eq. (24), we have,

$$\sqrt{\left(\Delta^2+k^2+(\varepsilon-\xi)^2\right)} = \frac{\cos^{-1}\left(\cos\left(\frac{\beta+\delta}{2}\right)\cos\left(\frac{\gamma}{2}\right)\right)}{2\pi T} \quad (30)$$

Equations (23), (28), (29) and (30) comprise four equations in five unknowns - $\Delta$, $\varepsilon$, $k$, $\xi$ and $T$. We therefore have the flexibility of choosing one of the parameters when solving for the unknowns to realize a desired controlled-unitary operation. Which parameter is chosen as the "fixed" parameter, depends on the physical implementation under consideration. For instance, for Josephson junction devices, the parameter $\Delta$, commonly called "tunneling", is fixed during the design phase.

As an example, suppose we wish to realize the CNOT gate. The **U** matrix in this case is the NOT gate. The values for the angles $\beta$, $\delta$ and $\gamma$ are $\pi$, 0 and $\pi$ respectively. Suppose we choose $\Delta$ to be 25 MHz, i.e., $\Delta_A = \Delta_B = 25$ MHz, which is typical for the tunneling parameter of SQUIDs [26]. Recall that since the value of $\varepsilon_A$ (which for SQUIDs constitutes the bias, in this case acting on the control qubit A) was chosen to be arbitrarily high (typically 10 GHZ in our simulations), we choose the value of the tunneling parameter of the control qubit to be the same as that of the target qubit. This is because as previously mentioned, under a large bias field, the effect of the tunneling parameter on the evolution of qubit A can be ignored. Substituting the values of $\beta$, $\delta$, $\gamma$, $\Delta$ and choosing $P = 1$ in Eqs. (23), (28), (29) and (30), we find the values of $\varepsilon$, $\xi$, $k$ and $T$ to be 48.4 MHz, 48.4 MHz, 0 and 10 ns respectively. Basically, during a CNOT gate operation, the coupling, $\xi$, between the target and control qubits is fixed at 48.4 MHz and then $\varepsilon_B$ (which for SQUID systems constitutes the bias acting on the target qubit) is pulsed from a high value (10 GHz) to a low value equal to 48.4 MHz for a time period of 10 ns. On simulating a two qubit coupled system using these parameters, we found that a CNOT gate is successfully realized, in agreement with our previous results presented in [26].

The main point of our scheme is to demonstrate improvement over conventional schemes that break up a controlled-unitary operation into a set of single qubit rotations and CNOT gates. To see this, suppose we wish to implement a controlled-H gate where H denotes the Hadamard gate [1]. For the **U** matrix to realize a Hadamard gate, one set of values for $\beta$, $\delta$ and $\gamma$ could be $2\pi$, $\pi$ and $\pi/2$, respectively. These values of the angles realize the Hadamard gate with an overall global phase shift of $\pi/2$. Once again, we choose

SQUIDs as our physical system and assuming the tunneling parameter is fixed during fabrication in our design, we use the same value of $\Delta$ as that used for the CNOT gate (25MHz) in finding the other system parameters. Substituting the values of $\beta$, $\delta$ and $\gamma$ with $\Delta =$ 25MHz and $P = 1$ in Eqs. (23), (28), (29) and (30), we find the values of $\varepsilon$, $\xi$ and $T$ to be 82.1MHz, 57.1MHz and 7ns, respectively (For a timestep of 10ns, the values of $\varepsilon$, $\xi$ and $\Delta$ to be 58.1MHz, 40.4MHz and 17.7MHz, respectively). To realize a controlled-Hadamard gate, the bias on the target qubit, $\varepsilon_B$, is simply pulsed to a lower value of 82.1 MHz for a time step of 7 ns. The gate operation is therefore realized in a single pulse, which is an advantage. Figure 4 shows how to realize a controlled-H gate (modulo a phase shift of $\pi/2$) using CNOT gates and single qubit rotations. Figure 5 show the actual pulse operations performed on the target qubit in order to realize the gate operation. Notice that each of the single qubit rotations require the coupling between the qubits to be turned off, necessitating the coupling, in addition to the bias, to be treated as a variable parameter of the system. For a single qubit system with Hamiltonian given by Eq. (1), arbitrary rotations are realized by varying the parameters, $\Delta$ and $\varepsilon$, of the qubit such that the unitary matrix given by Eq. (2) realizes the desired operation within a chosen time step.

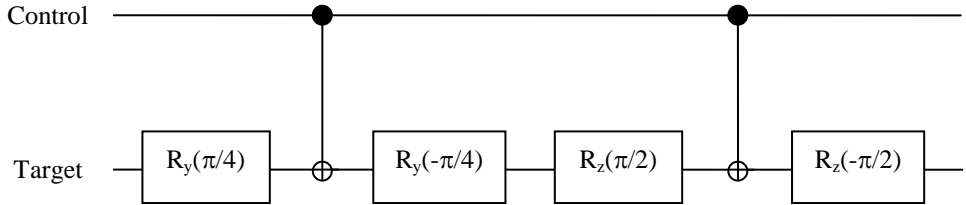

Fig. 4. Sequence of single-qubit and CNOT gate operations performed on the two-qubit system to realize a controlled-H gate using conventional methods. The gate is realized modulo an overall phase shift of $\pi/2$.

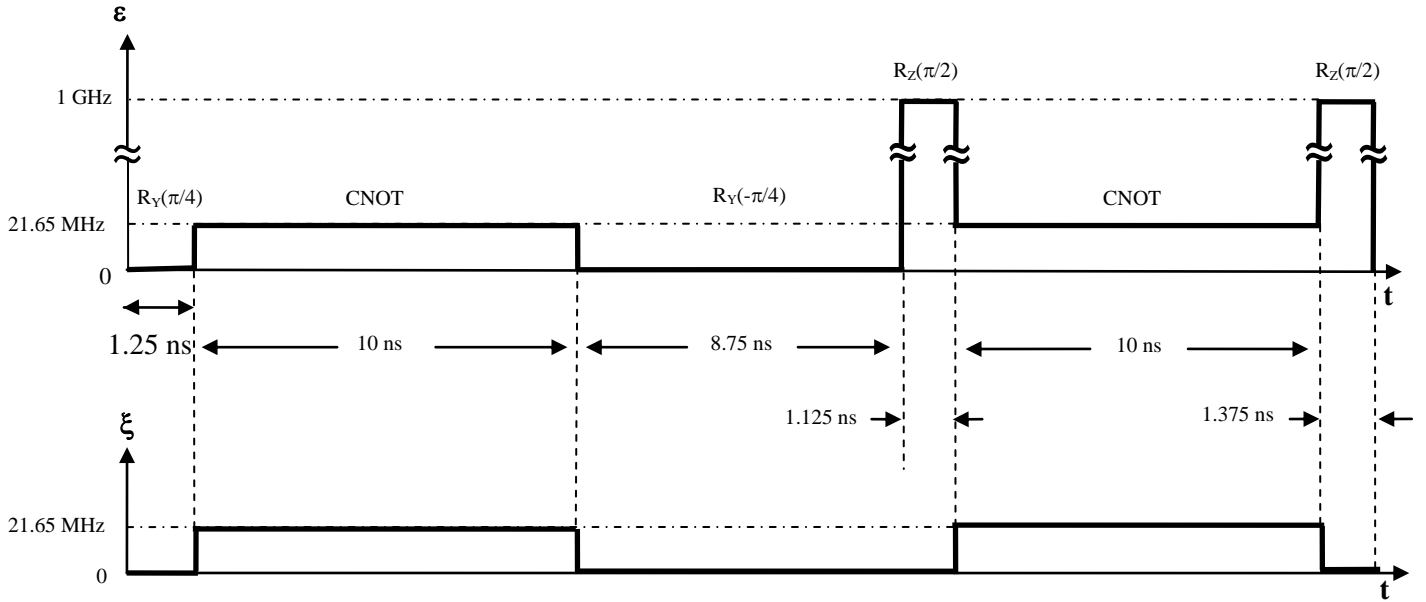

Fig. 5. Variation of the bias on the target qubit, $\varepsilon_B$, and the coupling, $\xi$, in order to realize the sequence of single-qubit and CNOT gate operations as shown in Fig. 4 to implement a controlled-H gate operation. The total time required to perform the gate operation is 32.5ns (Note that in realizing the single qubit rotation $R_Y(-\pi/4)$, we have used the fact that "$-\pi/4$" corresponds to "$7\pi/4$"). On the other hand, using our scheme, the controlled-H gate can be achieved in 7 ns when the tunneling parameter is chosen to be the same as that used in deriving parameters for the CNOT gate operation, i.e., $\Delta = 25$ MHz.. Moreover, the single qubit operations in the conventional scheme require the coupling to be a variable parameter of the system, which is a drawback when compared to our method which does not require this. The plots have not been drawn to scale due to space considerations.

As another example, suppose we wish to implement a controlled-unitary operation where the **U** matrix is given as:

$$\mathbf{U} = \mathbf{R}_Z(\beta)\mathbf{R}_Y(\gamma)\mathbf{R}_Z(\delta) = \mathbf{R}_Z\left(\frac{\pi}{2}\right)\mathbf{R}_Y\left(\frac{\pi}{2}\right)\mathbf{R}_Z\left(\frac{\pi}{3}\right) \tag{31}$$

In this case, we get a non-zero value for the parameter, $k$. By substituting the values of β, δ and γ, with $P = 1$ and $T = 10$ns, we find the values of the parameters $\varepsilon$, $\xi$, $\Delta$ and $k$ to be 57.0MHz, 41.7MHz, -4.12MHz and 15.3MHz, respectively. (We could have fixed the value of $\Delta$ at 25MHz and solved for the timestep instead). On simulating the gate operation under the unreduced Hamiltonian using these calculated parameters, the desired controlled-unitary operation is exactly realized. Table I shows the initial and final states of the two-qubit system under the controlled-unitary operation. Observing the probability amplitudes of the final states, we can see that the controlled-unitary operation is indeed exactly realized using the parameters calculated from our reduced Hamiltonian approach.

| Initial State | Final state |
|---|---|
| $\|00\rangle$ | $\|00\rangle$ |
| $\|01\rangle$ | $\|01\rangle$ |
| $\|10\rangle$ | $\frac{1}{\sqrt{2}}e^{-i75°}\|10\rangle + \frac{1}{\sqrt{2}}e^{i15°}\|11\rangle$ |
| $\|11\rangle$ | $\frac{1}{\sqrt{2}}e^{i165°}\|10\rangle + \frac{1}{\sqrt{2}}e^{i75°}\|11\rangle$ |

Table I. Simulation results showing the initial and final states of the system under the controlled-unitary operation where the **U** matrix is as given by Eq. (31). The parameters calculated using our reduced Hamiltonian approach were used in these simulations. Observing the probability amplitudes of the final states we can see that the desired controlled-unitary operation is exactly realized

It is important to point out that if the **U** matrix under consideration is diagonal, Eq. (29) will result in the condition $\Delta = 0$. Since $k$ is related to $\Delta$ by Eq. (28), we then have $k = 0$. In a physical system where we treat $\Delta$ as the variable parameter of the system, this can be achieved. However, as with the case of SQUIDs, if $\Delta$ is fixed during fabrication, it might be possible to make it zero as required to perform such gate operations. Under such cases, the parameters can be solved for by making the approximation that $\Delta^2 \ll 1/T^2$ (and therefore, $k^2 \ll 1/T^2$), for instance, we can choose $T$ such that $T = 50/\Delta$ where $\Delta = 25$ GHz as before. Using this approximation, the parameters $\varepsilon$ and $\xi$ can be calculated using Eqs. (23) and (30) by substituting for $T$ and neglecting $\Delta$ and $k$ in the calculations. The controlled-Z and controlled-phase gates are examples of gates where such an approximation is made.

The question arises whether our scheme can be generalized towards realizing controlled-unitary operations on a target qubit involving more than one control qubit. Suppose we have an "n+1" qubit system with the target qubit coupled through coupling terms, $\xi_1, \xi_2, \ldots, \xi_n$, to "n" controls. Such controlled-unitary operations are represented as $C^n(\mathbf{U})$ in the literature where the **U** matrix is as given by Eq. (20). We are assuming here that the control qubits do not interact with each other. As before, the states of the control qubits can be fixed by maintaining high values for the parameter $\varepsilon$ for each of these qubits. The $2^{n+1} \times 2^{n+1}$ Hamiltonian of the system can then be reduced to $2 \times 2$ Hamiltonians describing the evolution of the target qubit only. Since there are n controls, there will be $2^n$ Hamiltonians because the evolution of the target qubit takes place simultaneously in $2^n$ subspaces (Section III discussed the evolution for the case when n=1). Therefore, under a $C^n(\mathbf{U})$ operation if we were to write equations of the form of (23) through (27) for this system, we find that we always end up with "$2^n + 2$" equations in "n + 4" unknowns. This is because for all the cases where one or more of the control qubits are in the $|0\rangle$ state, we end up with "$2^n - 1$" equations of the form of Eq. (23). For the case where all the control qubits are in the $|1\rangle$ state, we end up with 3 equations similar to Eqs. (28) through (30), with the term "$\varepsilon - \xi$" replaced by the term "$\varepsilon - \xi_1 - \xi_2 - \xi_3 - \ldots - \xi_n$". Since the number of equations are greater than the number of unknowns, except for the cases when n=1 and n=2, the system of equations will thus be over-determined and it is, therefore, not possible to solve for the parameters. Thus, the reduced Hamiltonians approach cannot be extended to a more general case involving more than two control qubits. This is one of the pitfalls of the approach.

Next, consider the architectural layout scheme shown in Figure 6 where we each qubit is coupled only to those adjacent to it (qubits are represented as circles and the coupling between qubits as solid lines). Suppose we wish to perform a controlled-unitary operation between qubits A and X. Observe that since qubit X is coupled to qubits B, C and D, its evolution is controlled by the states of these qubits, which can be arbitrary quantum states. To isolate qubit X from the influence of these three qubits, it is required that we turn off the coupling between qubit X and qubits B, C and D, i.e., set $\xi_B$, $\xi_C$, $\xi_D$ to zero. However, in doing so, we are forced to treat the coupling as a variable of the system. One approach to overcome this is to fix the states of qubits B, C and D to the $|0\rangle$ state, in which case each of the three coupling terms, $\xi_B$, $\xi_C$, $\xi_D$ add to the bias term in the reduced Hamiltonian for qubit X, i.e., the reduced Hamiltonian for qubit X is of the form:

$$\mathbf{H_X} = \begin{pmatrix} \varepsilon_B \pm \xi_A + \xi_B + \xi_C + \xi_D & \Delta_B \\ \Delta_B & -(\varepsilon_B \pm \xi_A + \xi_B + \xi_C + \xi_D) \end{pmatrix} \quad (32)$$

Therefore, the same set of equations – (23), (28), (29) and (30) – can be used to realize a controlled-unitary operation with the terms ($\varepsilon$ +$\xi$) and ($\varepsilon$ -$\xi$) replaced by ($\varepsilon$ + $\xi_A$ + $\xi_B$ + $\xi_C$ + $\xi_D$) and ($\varepsilon$ - $\xi_A$ + $\xi_B$ + $\xi_C$ + $\xi_D$), respectively. The only drawback with this method is that it requires qubits B, C and D to be in the $|0\rangle$ state, which not always be easy to achieve.

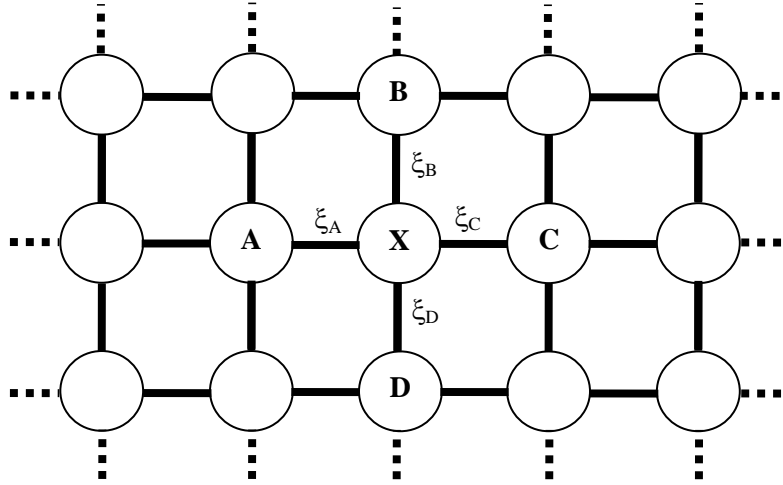

Fig. 6. A two-dimensional quantum computing architecture with nearest-neighbor interactions between qubits. Qubits are represented as circles and the coupling between qubits as solid lines. Here, we want to perform a controlled-unitary operation on qubit X without having to isolate qubit X from qubits A, B, C and D by switching off the coupling.

## V. Controlled-Unitary Operations with Heisenberg and XXZ Interactions

In the previous sections, we used the reduced Hamiltonian technique to find the parameters of a coupled system in order to realize controlled-unitary operations under an Ising interaction. While this coupling is used commonly to model the interaction between superconducting Josephson junction qubits, not all two-qubit systems interact through an Ising interaction. Isotropic coupling of qubits, also called Heisenberg interaction, characterized by coupling amplitudes $J_X = J_Y = J_Z$ is a very common form of two-body interaction that appears in many physical qubit proposals involving electronic spins, e.g., spin-coupled quantum dots [40, 41]. Equation (38) shows the most general form of the Hamiltonian of a two-qubit coupled system interacting via anisotropic interactions [42]:

$$\mathbf{H} = \begin{pmatrix} \varepsilon_A + \varepsilon_B + J_Z & \Delta_B & \Delta_A & J_X - J_Y \\ \Delta_B & \varepsilon_A - \varepsilon_B - J_Z & J_X + J_Y & \Delta_A \\ \Delta_A & J_X + J_Y & -\varepsilon_A + \varepsilon_B - J_Z & \Delta_B \\ J_X - J_Y & \Delta_A & \Delta_B & -\varepsilon_A - \varepsilon_B + J_Z \end{pmatrix}$$

(33)

Here, $J_X$, $J_Y$ and $J_Y$ are the coupling terms generated by the interaction Hamiltonian $J_X \; \sigma_X \cdot \sigma_X + J_Y \; \sigma_Y \cdot \sigma_Y + J_Z \; \sigma_Z \cdot \sigma_Z$. When $J_X = J_Y$, the interaction is the XXZ interaction. When $J_X = J_Y = J_Z$, the interaction is the Heisenberg interaction. When $J_Z = 0$, the interaction is called XY interaction. When $J_X = J_Y = 0$, the interaction is the familiar Ising interaction discussed in Section II. Here, $\sigma_X$, $\sigma_Y$, and $\sigma_Z$ are the Pauli spin matrices.

From Eq. (33) we can see that the reduced Hamiltonian approach used under an Ising interaction can be applied to the Hamiltonian matrix given by (33), provided that the coupling constants $J_X$ and $J_Y$ are much smaller than the parameter $\varepsilon_A$. We next check to see if our claim holds by simulating a particular controlled-unitary operation using the matrix given by Eq. (32) by using the same parameters values for which the gate operation was realized under the Ising interaction as described in Sections II and III. We choose the CNOT gate as the controlled-unitary operation, the value of the parameters for which is calculated as described in Section III under an Ising interaction. The parameters are: $\Delta_A = \Delta_B = 25$MHz, $\varepsilon_A = 10$GHz and $\varepsilon_B = 48.4$MHz. The value of the coupling constant $J_Z$ is chosen to be equal the coupling term, $\xi = 48.4$MHz, of the Ising interaction. The values of the coupling constants $J_X$ and $J_Y$ are arbitrarily chosen subject to the constraint that their values are much smaller than the parameter $\varepsilon_A$. Figure 7 shows the evolution of the probabilities of the four basis states under the Heisenberg interaction where $J_X = J_Y = J_Z = 48.4$MHz. As before, the bias on the target qubit is pulsed from a high value to a minimum value of 48.4MHz for a time duration of 10ns. From the plots, we can see that a CNOT gate operation is exactly realized. Therefore, as long as $J_X$ and $J_Y$ are much smaller than $\varepsilon_A$, the same parameters for which a CNOT gate operation was previously realized using an Ising interaction can also be used to realize the gate operation under anisotropic interactions, provided that the coupling constant $J_Z$ is equal to the value of the coupling parameter, $\xi$, derived under the Ising interaction.

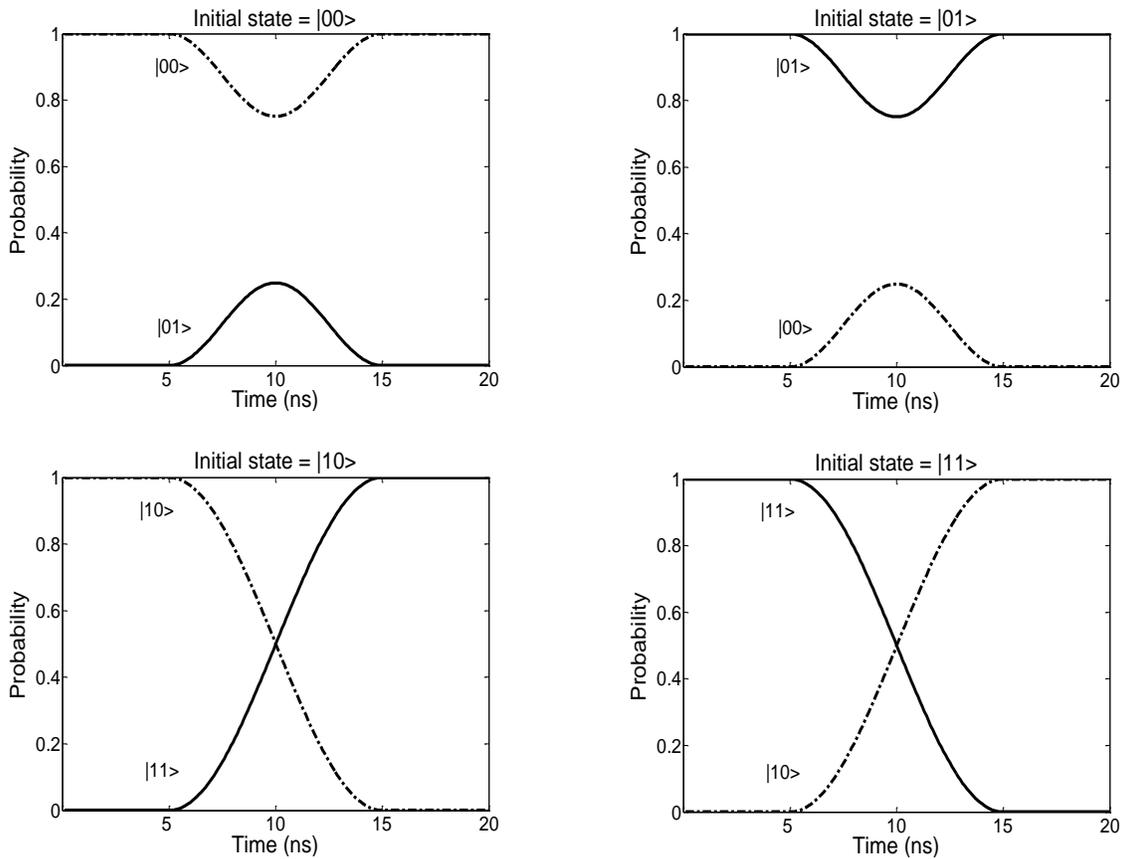

FIG 7. Evolution of the probabilities of the computational states under a CNOT gate operation when the initial state is each of the four basis states. The qubits interact through an Heisenberg interaction. The total time is 20ns. The parameter $\varepsilon_B$ of the target qubit is pulsed from a high value of 10 GHz to a low value of 48.4 MHz for a time duration of 10ns starting at 5ns. It is during this time interval that the transitions in the probabilities occur as can be seen from the plots.

## VI. Conclusions

We have shown here a means to realize arbitrary controlled-unitary operations in a two-qubit system. If the $2 \times 2$ unitary matrix belongs to the Lie group **SU(2)**, the gate operation can be realized in a *single pulse operation* where one of the parameters of the system is pulsed between an arbitrary high value and a calculated low value. We freeze the dynamics of the control qubit by controlling one of its parameters whereby the evolution of the two-qubit system reduces to that of a single qubit – the target. This allows us to reduce the coupled system Hamiltonian to a $2 \times 2$ matrix describing the evolution of the target qubit only. There are two different Hamiltonians describing the evolution of the target qubit depending upon the state of the control qubit. Using these reduced Hamiltonians, we calculate the parameters of the two-qubit system such that the desired controlled-unitary gate operation is achieved up to an overall global phase shift. The coupling parameter is treated as a fixed parameter of the system. We are, thus, able to realize a controlled-unitary operation on a system of two-qubits without having to switch off the coupling. Moreover, our scheme can be applied to a wide range of coupling schemes and can be used to realize gate operations between two qubits coupled via Ising, Heisenberg, XXZ and anisotropic interactions. The main advantage of our scheme is that the computational complexity for realizing an arbitrary two-qubit operation in a general two-qubit system is equivalent to realizing a CNOT gate. Since conventional schemes realize a controlled-unitary operation by breaking it into a sequence of single-qubit and CNOT gate operations, our method is an improvement because we not only require lesser time duration, but also fewer control lines, to implement the same operation.

________________________________________